\documentclass[aps,prb,twocolumn,superscriptaddress]{revtex4}

\usepackage{graphicx}% Include figure files

\begin{document}

%\preprint{}

\title{ Photoinduced melting of charge order in a quarter-filled electron system coupled with different types of phonons }

% Please use the appropriate macro for each type of information

\author{ Kenji Yonemitsu }
\email[]{kxy@ims.ac.jp}
%\thanks{}
\affiliation{Institute for Molecular Science, Okazaki 444-8585, Japan}
\affiliation{Department of Functional Molecular Science, Graduate University for Advanced Studies, Okazaki 444-8585, Japan}
\author{ Nobuya Maeshima }
%\email[]{}
%\thanks{}
\affiliation{Institute for Molecular Science, Okazaki 444-8585, Japan}
\affiliation{Department of Chemistry, Tohoku University, Aramaki, Aoba-ku, Sendai 980-8578, Japan}

\date{\today}

\begin{abstract}
Photoinduced melting of charge order is calculated by using the exact many-electron wave function coupled with classically treated phonons in the one-dimensional quarter-filled Hubbard model with Peierls and Holstein types of electron-phonon couplings. The model parameters are taken from recent experiments on (EDO-TTF)$_2$PF$_6$ (EDO-TTF=ethylenedioxy-tetrathiafulvalene) with (0110) charge order, where transfer integrals are modulated by molecular displacements (bond-coupled phonons) and site energies by molecular deformations (charge-coupled phonons). The charge-transfer photoexcitation from (0110) to (0200) configurations and that from (0110) to (1010) configurations have different energies. The corresponding excited states have different shapes of adiabatic potentials as a function of these two phonon amplitudes. The adiabatic potentials are shown to be useful in understanding differences in the photoinduced charge dynamics and the efficiency of melting, which depend not only on the excitation energy but also on the relative phonon frequency of the bond- and charge-coupled phonons. 
\end{abstract}

% 78.20.Bh Theory, models, and numerical simulation  
% 71.45.Lr Charge-density-wave systems (see also 75.30.Fv Spin-density waves) 
% 71.10.Fd Lattice fermion models (Hubbard model, etc.)  
% 63.20.Kr Phonon-electron and phonon-phonon interactions  
% 71.30.+h Metal-insulator transitions and other electronic transitions  
% 71.38.-k Polarons and electron-phonon interactions (see also 63.20.Kr)
% 71.45.-d Collective effects
% 78.47.+p Time-resolved optical spectroscopies and other ultrafast optical measure.

\maketitle

\section{Introduction}

Introduction of electrons and holes by photoirradiation of an insulator may trigger a phase transition, called a photoinduced phase transition, which are often characterized by a nonlinear relation between the field and the response. The electronic and structural states evolve in a cooperative manner, leading to macroscopic changes in the physical properties. \cite{nasu_book04,jpsj_photoinduced06} They may create novel transient electronic states or dynamical functions. Thus, it is important to search for possibility of controlling photoinduced phase transition dynamics. 

When an ordered state is stabilized by different types of interactions, e.g., by electron-electron (e-e) and electron-phonon (e-ph) interactions or by different types of e-ph interactions, their relative contributions would be clarified by investigating the photoinduced dynamics of the electron-phonon state. For instance, when an e-ph interaction stabilizes the order, the motion of photoinduced charge transfers and that of phonons are strongly coupled and often show a coherent oscillation in transient reflectivity. \cite{iwai_prl02,cavalleri_prb04,chollet_s05,okamoto_prl06}

In most cases, one type of phonons has been considered in discussing the stabilization of an ordered state and its photoinduced dynamics. \cite{yonemitsu_prb03,yonemitsu_prb06} If different types of phonons contribute to forming a charge order by modulating different electronic parameters, complex dynamics are expected and their analysis would give valuable information that cannot be easily obtained by measuring the physical properties in thermal equilibrium. 

In this paper, we consider the organic salt (EDO-TTF)$_2$PF$_6$ with 3/4-filled band (1/4-filled in terms of holes), which shows a metal-insulator transition with (0110) charge order accompanied by a large structural change based on a doubling of unit cells and deformation of EDO-TTF molecules at low temperatures. \cite{ota_jmc02} The (0110) charge order is thus stabilized by two different types of phonons, transfer-integral-modulating displacements (bond-coupled phonons) and site-energy-modulating deformations (charge-coupled phonons). An ultrafast transition (within 3 ps) from the charge-ordered insulator phase to the ``metal'' phase is induced by weak laser light with a pulse width of 0.12 ps and of excitation photon energy 1.55 eV \cite{chollet_s05} nearly resonant with charge transfer toward the (0200) configuration. \cite{drozdova_prb04} The observed spectroscopic change shows cooperative melting of the charge order assisted by coherent phonon generation. \cite{chollet_s05,onda_jpcs05}

For quarter-filled systems, different charge-transfer excited states generally have different shapes of adiabatic potentials and thus different directions of initial force applied to phonon amplitudes. The photoinduced dynamics would have respective efficiency, and some of them can be sensitive to the relative phonon frequency, as shown in this paper. This fact may be useful in identifying the phonon responsible for the coherent oscillation, and more generally, in designing a material with more efficient transition dynamics. 

\section{Peierls-Holstein-Hubbard Model}

As a model for (EDO-TTF)$_2$PF$_6$, we treat the one-dimensional quarter-filled Hubbard model with Peierls and Holstein types of e-ph couplings, which modulate transfer integrals and site energies, respectively. 
\begin{eqnarray}
H & = & -\sum_{j,\sigma} 
\left[ t_0 -\alpha \left( u_{j+1} - u_{j} \right) \right]
( c^\dagger_{j,\sigma} c_{j+1,\sigma} + c^\dagger_{j+1,\sigma} c_{j,\sigma} )
\nonumber \\ & &
-\beta \sum_j v_j ( n_j -\frac12 ) + U \sum_j n_{j,\uparrow} n_{j,\downarrow}
\nonumber \\ & &
+\frac12 K_\alpha \sum_j \left( u_{j+1} - u_{j} \right)^2 
+\frac12 K_\beta  \sum_j v_j^2 
\nonumber \\ & & 
+\frac{2K_\alpha}{\omega_\alpha^2} \sum_j \dot{u}_j^2 
+\frac{K_\beta}{2\omega_\beta^2} \sum_j \dot{v}_j^2 
\;,
\end{eqnarray}
where $ c^{\dagger}_{j,\sigma} $ ($ c_{j,\sigma} $) is the creation (annihilation) operator of a hole with spin $\sigma$ at site $j$, $ n_{j,\sigma} = c^{\dagger}_{j,\sigma} c_{j,\sigma} $, $ n_{j} = n_{j,\uparrow} + n_{j,\downarrow} $, $ u_j $ is the displacement of the $ j $th molecule from equilibrium, $ v_j $ is its deformation amplitude, and $ \dot{u}_j $ and $ \dot{v}_j $ are the time derivatives of $ u_j $ and $ v_j $, respectively. The band-filling is a quarter in terms of holes. The parameter $ t_0 $ denotes the bare transfer integral, which is assumed to be uniform for simplicity, $ U $ the on-site repulsion strength, $ \alpha $ and $ \beta $ the e-ph coupling strengths of modulating the transfer integrals by the molecular displacements (bond-coupled phonons) and the site energies by the molecular deformations (charge-coupled phonons), respectively, $ K_\alpha $ and $ K_\beta $ the corresponding spring constants, and $ \omega_\alpha $ and $ \omega_\beta $ the corresponding bare phonon frequencies. In reality, the transfer integrals would be modulated by the molecular deformations as well, but such details are ignored. The static properties of this model ($ \omega_\alpha $=$ \omega_\beta $=0) have already been studied so intensively, including exact-diagonalization \cite{mazumdar_prb00,clay_prb03} and density-matrix-renormalization-group \cite{kuwabara_jpsj03} studies. For charge orders in quarter-filled organic compounds, see for example Ref.~\onlinecite{seo_jpsj06}.

\begin{figure}
\includegraphics[height=13cm]{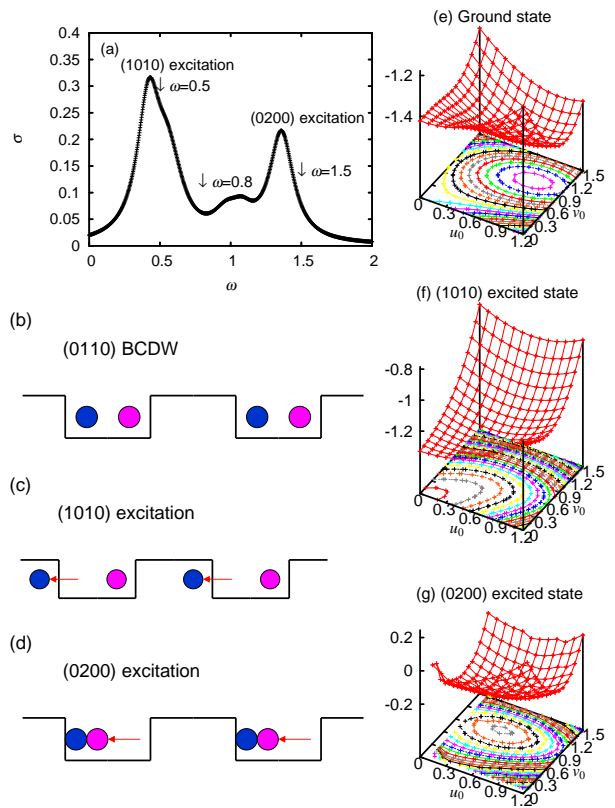}
\caption{(Color online) (a) Optical conductivity in (0110) ground state with 12 sites. Schematic charge distribution (circles) and deformation-induced potential (line), for (b) (0110) ground state, (c) (1010) photoexcited state, and (d) (0200) photoexcited state. (e), (f), and (g) are their adiabatic potentials as a function of amplitudes of bond-($ u_0 $) and charge-($ v_0 $)coupled phonons. 
\label{fig:opt_cg_adpot}}
\end{figure}

In the (EDO-TTF)$_2$PF$_6$ salt, displacements of anions in the insulator phase are substantial \cite{aoyagi_acie04} and favor the (0110) (called bond-charge-density wave in Ref.~\onlinecite{clay_prb03}) ground state  because the neighboring site energies are modulated in phase. This effect of anions counterbalances that of the nearest-neighbor repulsion favoring the (1010) (called 4$ k_F $ charge-density wave spin-Peierls in Ref.~\onlinecite{clay_prb03}) ground state. We here ignore such details for simplicity and assume that the effect of anions would renormalize the parameters related to the charge-coupled phonons, $ \beta $, $ K_\beta $, and $ \omega_\beta $. Then, most of the model parameters can be deduced from the charge disproportionation and the optical conductivity spectrum, as performed in Ref.~\onlinecite{drozdova_prb04}: $ t_0 $=0.16, $ U $=0.93, $ \alpha^2/K_\alpha $=0.14, and $ \beta^2/K_\beta $=0.55, in units of eV. The classically treated, bond- and charge-coupled phonons are obtained by imposing the Hellmann-Feynman theorem. The optical conductivity spectrum is then obtained with peak-broadening parameter set at 0.1 and shown in Fig.~\ref{fig:opt_cg_adpot}(a). 
Because the on-site repulsion $ U $ is assumed to be the largest energy, the high-energy peak corresponds to the charge-transfer excitation to the (0200) configuration ($ \omega_{0200} \sim $1.4), while the low-energy peak to that to the (1010) configuration ($ \omega_{1010} \sim $0.4). 

\section{Adiabatic Potentials}

The spring constants $ K_\alpha $ and $ K_\beta $ are so chosen that $ \mid u_j \mid $ and $ \mid v_j \mid $ are of the order of unity: $ K_\alpha $=0.035 and $ K_\beta $=0.1 to give $ u_j $=0.59, 0.85, $-$0.85, $-$0.59, $ \cdots $ and $ v_j $=$-$1.01, 1.01, 1.01, $-$1.01, $ \cdots $ for the 12-site periodic chain. The charge density per site is 0.07, 0.93, 0.93, 0.07, $ \cdots $ in the ground state, though it is simply denoted by (0110). The charge distribution $ \langle n_j \rangle $ and the deformation-induced potential $ -\beta v_j $ are schematically shown by the circles and the line, respectively, in Figs.~\ref{fig:opt_cg_adpot}(b)-\ref{fig:opt_cg_adpot}(d) for the ground and photoexcited states. The corresponding adiabatic potentials are plotted in Figs.~\ref{fig:opt_cg_adpot}(e)-\ref{fig:opt_cg_adpot}(g). They are calculated with varying $ u_j $ and $ v_j $ through similar extension/contraction from the optimized configurations mentioned above. The maximum values of $ u_j $ and $ v_j $ are denoted by $ u_0 $ and $ v_0 $, respectively. 

The minimum in the adiabatic potential of the (1010) photoexcited state is located at the origin [Fig.~\ref{fig:opt_cg_adpot}(f)]. The excitation energy is lowered by delocalizing the photocarriers, so that both of $ u_j $ and $ v_j $ are strongly suppressed. In contrast, the minimum in the adiabatic potential of the (0200) photoexcited state is shifted from that of the ground state to reduce $ u_j $ only [Fig.~\ref{fig:opt_cg_adpot}(g)]. The energy contribution from the deformation-induced potential is basically unchanged by the charge-transfer process shown in Fig.~\ref{fig:opt_cg_adpot}(d). Meanwhile, the displacements shortening the distance between the neighboring hole-rich sites and stabilizing the tetramers are relaxed. Therefore, the shift of the stable phonon configuration is quite anisotropic by the (0200) photoexcitation. 

\section{Photoinduced Melting Dynamics}

\subsection{Photoexcitation by an oscillating field}

Photoexcitations are introduced by adding the modified transfer term with the Peierls phase and the coupling with the molecular displacements (bond-coupled phonons) to the Hamiltonian, 
\begin{eqnarray}
& & -\sum_{j,\sigma} 
\left[ t_0 -\alpha \left( u_{j+1} - u_{j} \right) \right]
\left\{ \exp \left[ \frac{iea}{\hbar} \int dt E(t) \right] -1 
\right\} \nonumber \\ & & \times
c^\dagger_{j,\sigma} c_{j+1,\sigma} + \text{h.c.} 
-\sum_j e \langle n_j \rangle 
\delta u_j E(t)
\;, \label{eq:electric_field}
\end{eqnarray}
where $ e $ is the absolute value of the electronic charge, $ a $ is the lattice spacing, $ \delta $=0.05$ a $ is introduced to recover the length dimension. The time-dependent electric field $ E(t) $ is given by 
$ E(t) = -E_{\text{ext}} \sin \omega_{\text{ext}} t $ 
with amplitude $ E_{\text{ext}} $ and frequency $ \omega_{\text{ext}} $ for $ 0 < t < T_{\text{irr}} $ [$ E(t) $ is zero otherwise], where the pulse width is set at $ T_{\text{irr}} = 2 \pi N_{\text{ext}}/\omega_{\text{ext}} $ with integer $ N_{\text{ext}} $. 

It is easily shown by the gauge transformation that the vector potential introduced above is equivalent to the scalar potential. \cite{yonemitsu_jpsj05} The vector potential is convenient because the boundary condition remains periodic. Of the two terms in Eq.~(\ref{eq:electric_field}), the first is much larger than the second. Even if the second term is omitted, the numerical results below are almost unchanged. The pulse width $ T_{\text{irr}} $ is set to nearly coincide with the experimentally used value of about 200 in the present unit. The time-dependent Schr\"odinger equation for the exact many-electron wave function is numerically solved by expanding the exponential evolution operator with time slice $ dt $=0.01 to the 15th order and by checking the conservation of the norm and of the total energy for $ t > T_{\text{irr}} $. The classical equation of phonon motion is solved by the leapfrog method, where the force is derived from the Hellmann-Feynman theorem. The phonon frequencies $ \omega_\alpha $ and $ \omega_\beta $ are varied in such a way that one of the renormalized frequencies is close to the experimentally observed value of about 0.01. (either $ \omega_\alpha $ or $ \omega_\beta $ is set to be 0.02.)

\subsection{Numerical results}

First, the (0110) ground state is excited with $ \omega_{\text{ext}} $=1.5 and $ N_{\text{ext}} $=50. 
\begin{figure}
\includegraphics[height=16cm]{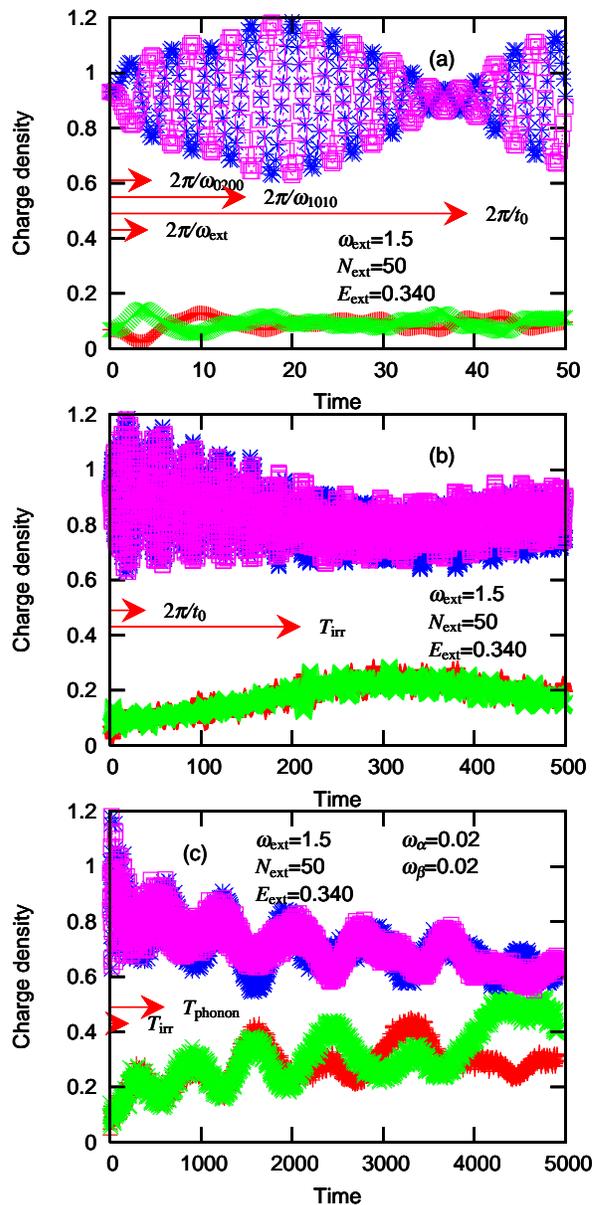}
\caption{(Color online) Time dependence of charge densities, (a), (b), and (c) on short, intermediate, and long time scales, respectively, for excitation energy $ \omega_{\text{ext}} $=1.5 close to $ \omega_{0200} $, $ N_{\text{ext}} $=50, and $ E_{\text{ext}} $=0.340 just below threshold level, for $ \omega_\alpha $=$ \omega_\beta $=0.02. 
\label{fig:dnst_evol}}
\end{figure}
The evolution of charge densities at an early stage is shown in Fig.~\ref{fig:dnst_evol}(a). The period of $ E(t) $ is $ 2\pi/\omega_{\text{ext}} \simeq $4.2, which is close to that of the (0200) excitation, $ 2\pi/\omega_{0200} \simeq $4.6, and far from that of the (1010) excitation, $ 2\pi/\omega_{1010} \simeq $15. The bare transfer integral has the time scale of $ 2\pi/t_0 \simeq $39. Because of $ \omega_{\text{ext}} \sim \omega_{0200} $, the charge densities at the hole-rich sites are largely oscillating with period $ 2\pi/\omega_{\text{ext}} $, like (0110)$ \rightarrow $(0200)$ \rightarrow $(0110)$ \rightarrow $(0020). The oscillation at the hole-poor sites is initially small. This forced oscillation mainly at the hole-rich sites continues until the field is turned off at $ t $=$ T_{\text{irr}} \simeq $210 [Fig.~\ref{fig:dnst_evol}(b)]. By this time, the charge-density difference between the hole-rich and the hole-poor sites is reduced by the weakened $ u_j $ and $ v_j $. Even for $ t > T_{\text{irr}} $, the rapid oscillations due to charge-transfer excitations are evident. In addition, the slow oscillation on the phonon time scale, $ T_{\text{phonon}} \simeq $600, is clearly seen in Fig.~\ref{fig:dnst_evol}(c). This coherent oscillation is allowed by the strong coupling between the charge and phonon dynamics. Here, $ E_{\text{ext}} $ is chosen to be just below the threshold for melting. Then, after several periods of the phonon oscillation, the charge configuration is nearly disordered and the charge order is quite reduced. 

\begin{figure}
\includegraphics[height=16cm]{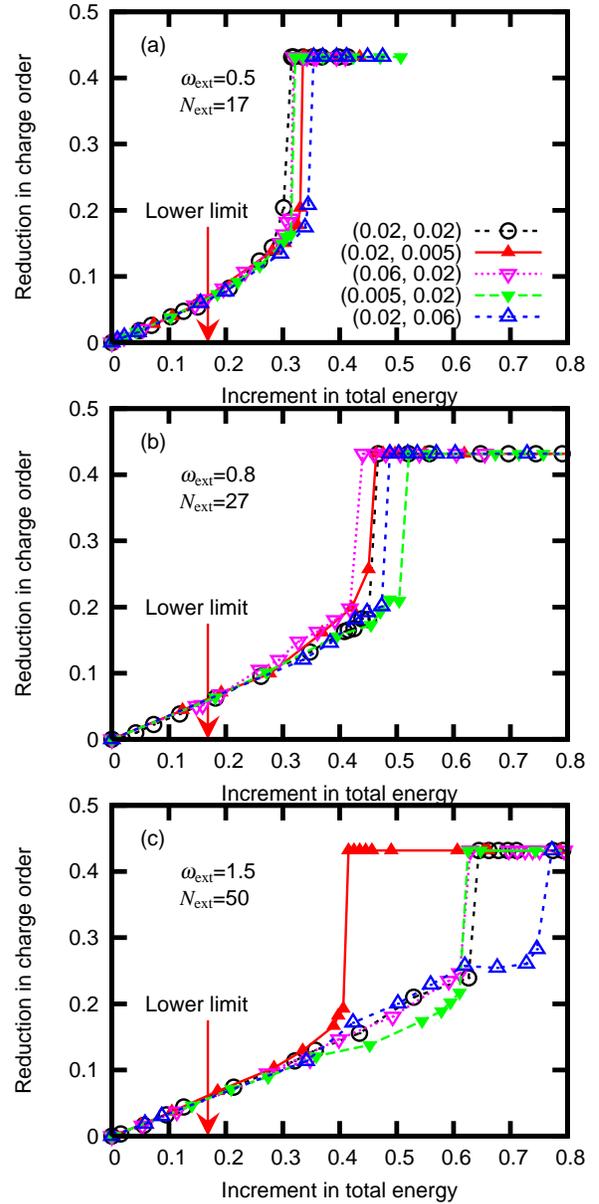}
\caption{(Color online) Reduction in charge order averaged over the period of oscillation at $ t \leq $4000, as a function of increment in total energy, for 
(a) $ \omega_{\text{ext}} $=0.5, $ N_{\text{ext}} $=17, 
(b) $ \omega_{\text{ext}} $=0.8, $ N_{\text{ext}} $=27, and 
(c) $ \omega_{\text{ext}} $=1.5, $ N_{\text{ext}} $=50. 
Bare phonon frequencies are ($ \omega_\alpha $, $ \omega_\beta $)=(0.02, 0.02) (black), (0.02, 0.005) (red), (0.06, 0.02) (magenta), (0.005, 0.02) (green), and (0.02, 0.06) (blue).
The arrows show the lower limit required for the melting explained in the text. 
\label{fig:red_chg_mod}}
\end{figure}

How much the charge order is reduced can be quantified by averaging the charge density at the (initially) hole-poor sites over the period of the phonon oscillation at $ t \leq $4000 (corresponding to an interval inside 2 to 3 ps) and by subtracting the initial density from it. It is plotted in Fig.~\ref{fig:red_chg_mod} as a function of the increment in the total energy $ \Delta E $, for different excitation energies $ \omega_{\text{ext}} $ [indicated by the arrows in Fig.~\ref{fig:opt_cg_adpot} (a)] and different phonon frequencies $ \omega_\alpha $ and $ \omega_\beta $. 
The arrows here show the energy difference between the undistorted and the ground states, which gives the lower limit of energy required for the melting. In all cases, the responses are linear for small $ \Delta E $, and their linear coefficients are nearly the same. For substantial $ \Delta E $, the curves are convex, indicating cooperativity in the photoinduced dynamics. It should be noted that, even if the charge order is almost completely destroyed by averaging over the phonon oscillation period, snapshots of charge densities and amplitudes of bond- and charge-coupled phonons show quite disproportionate configurations. Thus, the photoinduced melting does not produce a regular metallic phase. 

For excitations of energy $ \omega_{\text{ext}} $=0.5 close to $ \omega_{1010} $, the curve is insensitive to the combination of phonon frequencies  [Fig.~\ref{fig:red_chg_mod}(a)]. Compared with the other excitation energies discussed below, the charge order is melted by smaller $ \Delta E $ (0.31$ < \Delta E < $0.35). In other words, the efficiency of photoinduced melting is the highest on average. This fact is similar to the fact, in one-dimensional half-filled dimerized Mott insulators, interdimer charge-transfer excitations reduce the spin-Peierls order much more efficiently than intradimer ones, leading to a photoinduced inverse spin-Peierls transition. \cite{maeshima_prb06} In the present quarter-filled case also, carriers photogenerated by (1010) charge-transfer excitations are delocalized, so that the adiabatic potential has a minimum at the undistorted phonon configuration [Fig.~\ref{fig:opt_cg_adpot}(f)] to maximize the kinetic energy gain. 

For excitations of energy $ \omega_{\text{ext}} $=0.8 on the high-energy side of the (1010) charge-transfer peak, the efficiency of photoinduced melting is lowered and depends weakly on the phonon frequencies [Fig.~\ref{fig:red_chg_mod}(b)]. The lower efficiency than the case with $ \omega_{\text{ext}} $ closer to $ \omega_{1010} $ is reasonable in the sense that off-resonant excitations contain processes not simply represented by the (0110)$ \rightarrow $(1010) charge transfer: extra energy is needed to transfer charge. The dependence of the efficiency on the phonon frequencies has some regularity. Here, the curves for $ \omega_\alpha > \omega_\beta $ are plotted in warm colors, while those for $ \omega_\alpha < \omega_\beta $ in cold colors. The former case generally has a higher efficiency than the latter. 

This fact is more conspicuous for excitations of energy $ \omega_{\text{ext}} $=1.5 close to $ \omega_{0200} $ [Fig.~\ref{fig:red_chg_mod}(c)]. Among the three excitation energies, the charge order is melted by the largest $ \Delta E $ on average, indicating the lowest efficiency. However, the curve strongly depends on the phonon frequencies, so that the average efficiency is not so meaningful. In fact, for ($ \omega_\alpha $, $ \omega_\beta $)=(0.02, 0.005), the efficiency is higher than the corresponding case of $ \omega_{\text{ext}} $=0.8. Meanwhile, the efficiency for ($ \omega_\alpha $, $ \omega_\beta $)=(0.02, 0.06) is very low. The origin of such dependence on the phonon frequencies is discussed below on the basis of the adiabatic potential, although the phonons do not necessarily evolve in its steepest descent direction. 

As mentioned before, the adiabatic potential of the (0200) photoexcited state [Fig.~\ref{fig:opt_cg_adpot}(g)] has a minimum shifted from that of the ground state [Fig.~\ref{fig:opt_cg_adpot}(e)] to reduce $ u_j $ only. Thus, the initial force reduces $ u_j $. For $ \omega_\alpha > \omega_\beta $, the energy supplied from the field is first transferred to the faster $ u_j $ motion, and then to the slower $ v_j $ motion, reducing both phonon amplitudes and the charge order in an efficient manner. Otherwise, the initially induced $ u_j $ motion is slow, and the secondary $ v_j $ motion requires higher energy than the first one, which is inefficient. 

The experiment on (EDO-TTF)$_2$PF$_6$ shows a correspondence of the frequency of the coherent oscillation with that of a Raman-active optical mode, \cite{chollet_s05} but its mode assignment is not done yet. This is why we vary the phonon frequencies in the calculations. At first glance, it may seem reasonable to assume that the displacement ($ u_j $) is generally more collective and has a lower frequency $ \omega_\alpha $ than the deformation ($ v_j $) regarded as an intramolecular mode. However, the situation is not so simple because the displacements of anions contribute to stabilize the (0110) charge order, as mentioned earlier. 

In order to study the effects of anion displacements and nearest-neighbor repulsion on the phase diagram, we add to the model, 
\begin{equation}
-\gamma \sum_l w_l ( n_{2l-1} + n_{2l} - 1 )
+\frac12 K_\gamma  \sum_l w_l^2 
+V \sum_j n_{j} n_{j+1}
\;,
\end{equation}
where $ w_l $ is the displacement of the $ l $th anion from equilibrium, $ \gamma $ the corresponding e-ph coupling strength, $ K_\gamma $ the corresponding spring constant, and $ V $ the nearest-neighbor repulsion strength. We use the relations $ U - V $=0.93 and $ \beta^2/K_\beta + 2\gamma^2/K_\gamma $=0.55 so as to minimize the modification of the optical conductivity spectrum, although $ V $ still sharpens and heightens the lowest-energy peak. The phase diagram is divided into the present (0110) order approximately for $ V < 0.1 + \gamma/1.3 $ and the (1010) order for $ V > 0.1 + \gamma/1.3 $. The almost linear relation is due to the competition between $ \gamma $ favoring in-phase modulation of $ \langle n_{2l-1} \rangle $ and $ \langle n_{2l} \rangle $ and $ V $ favoring out-of-phase modulation of them. The strong reduction of the critical $ V $ is due to the strong Holstein coupling $ \beta^2/K_\beta $. Without e-ph couplings ($ \alpha $=$ \beta $=0), the (1010) order needs $ V > 2t_0 $ in the limit of large $ U $ and a somewhat larger $ V $ for finite $ U $. Therefore, without anion displacements ($ \gamma $=0), the nearest-neighbor repulsion of only 10\% of the magnitude of $ U $ is thus found to convert the (0110) order into the (1010) order. Because the anion displacements strengthen the e-ph coupling $ \beta $, they make the $ v_j $ phonon effectively heavier. Thus, it is possible that the $ v_j $ phonon has a lower frequency $ \omega_\beta $. If it is the case, the efficiency with $ \omega_{\text{ext}} $=1.5 can be higher than that with $ \omega_{\text{ext}} $=0.8. It should be noted that, if the reduction in the charge order is plotted as a function of the number of absorbed photons ($ \Delta E / \omega_{\text{ext}} $), the excitations with $ \omega_{\text{ext}} $=1.5 are the most efficient and those with $ \omega_{\text{ext}} $=0.5 are the least efficient on average. 

Details of the actual dynamics of charge densities and phonon amplitudes are more complicated than those naively expected from the adiabatic potential as described above. The charge dynamics for $ \omega_\alpha \simeq \omega_\beta $ were close to a sinusoidal evolution [Fig.~\ref{fig:dnst_evol}(c)], but they are not always the case. Below we will describe the actual dynamics for $ \omega_\alpha \neq \omega_\beta $. 

\begin{figure}
\includegraphics[height=13cm]{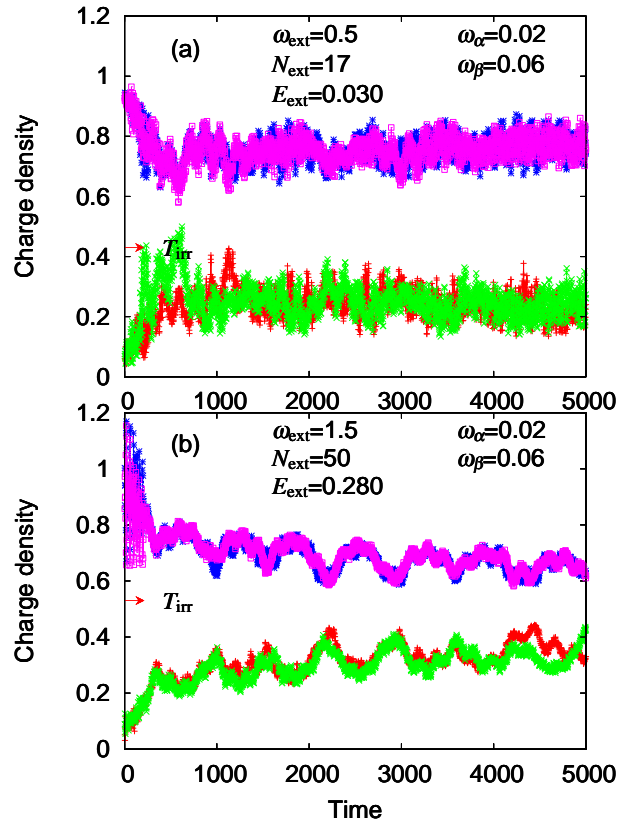}
\caption{(Color online) Time dependence of charge densities for $ \omega_\alpha $=0.02 and $ \omega_\beta $=0.06. Photoexcitations are (a) $ \omega_{\text{ext}} $=0.5 close to $ \omega_{1010} $, $ N_{\text{ext}} $=17, and $ E_{\text{ext}} $=0.030; and (b) $ \omega_{\text{ext}} $=1.5 close to $ \omega_{0200}, N_{\text{ext}} $=50, and $ E_{\text{ext}} $=0.280. 
\label{fig:dnst_evol_w2w6}}
\end{figure}

For $ \omega_\alpha < \omega_\beta $, the charge dynamics sensitively depend on the excitation energy (Fig.~\ref{fig:dnst_evol_w2w6}). 
When the excitation energy is close to $ \omega_{1010} $, $ \omega_{\text{ext}} $=0.5, the main component in the charge-density oscillation is due to the $ v_j $ phonon with the higher energy [Fig.~\ref{fig:dnst_evol_w2w6}(a)]. Because the corresponding adiabatic potential is rather isotropic with respect to the $ u_j $ and $ v_j $ amplitudes [Fig.~\ref{fig:opt_cg_adpot}(f)], the phonon with a stronger coupling strength ($ \beta^2/K_\beta > \alpha^2/K_\alpha $), $ v_j $, more contributes to the charge-density oscillation. On the other hand, when the excitation energy is close to $ \omega_{0200} $, $ \omega_{\text{ext}} $=1.5, the main component in the charge-density oscillation is due to the $ u_j $ phonon with the lower energy [Fig.~\ref{fig:dnst_evol_w2w6}(b)]. This is because the minimum in the corresponding adiabatic potential is shifted from that of the ground state to reduce $ u_j $ only [Fig.~\ref{fig:opt_cg_adpot}(g)]. 

\begin{figure}
\includegraphics[height=13cm]{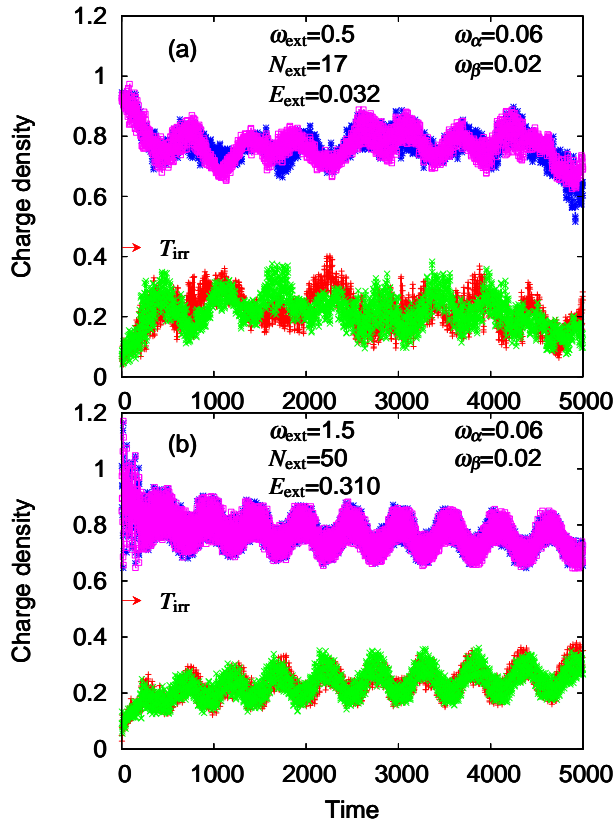}
\caption{(Color online) Time dependence of charge densities for $ \omega_\alpha $=0.06 and $ \omega_\beta $=0.02. Photoexcitations are (a) $ \omega_{\text{ext}} $=0.5 close to $ \omega_{1010} $, $ N_{\text{ext}} $=17, and $ E_{\text{ext}} $=0.032; and (b) $ \omega_{\text{ext}} $=1.5 close to $ \omega_{0200}, N_{\text{ext}} $=50, and $ E_{\text{ext}} $=0.310. 
\label{fig:dnst_evol_w6w2}}
\end{figure}

For $ \omega_\alpha > \omega_\beta $, the charge dynamics are much more insensitive to the excitation energy (Fig.~\ref{fig:dnst_evol_w6w2}) than for $ \omega_\alpha < \omega_\beta $ described above. 
The charge-density oscillation is dominated by a single component, whose frequency weakly depends on $ \omega_{\text{ext}} $. The frequency of this coherent oscillation for $ \omega_{\text{ext}} $=0.5 [Fig.~\ref{fig:dnst_evol_w6w2}(a)] is slightly lower than that for $ \omega_{\text{ext}} $=1.5 [Fig.~\ref{fig:dnst_evol_w6w2}(b)]. This small difference would be caused by the difference in the coupling strengths, $ \beta^2/K_\beta > \alpha^2/K_\alpha $, and the difference in the shapes of the adiabatic potentials, as in the case of $ \omega_\alpha < \omega_\beta $. The charge dynamics for ($ \omega_\alpha $, $ \omega_\beta $)=(0.02, 0.005) are slower but otherwise similar to the present case for ($ \omega_\alpha $, $ \omega_\beta $)=(0.06, 0.02). Such ($ \omega_\alpha $, $ \omega_\beta $)-dependent theoretical dynamics would be helpful in assigning experimentally observed oscillations. 

\section{Summary}

Photoinduced melting of charge order is calculated exactly for electrons and classically for phonons in the one-dimensional quarter-filled Hubbard model with different types of e-ph couplings. The strengths of the e-e and e-ph interactions and the transfer integral are taken to reproduce the charge disproportionation and the optical conductivity in the (0110) ground state of (EDO-TTF)$_2$PF$_6$. The efficiency of (0200) photoexcitations is generally (but not always) lower than that of (1010) photoexcitations, which generate delocalized carriers. Relative dynamics of the two phonon amplitudes are influenced by the shape of the adiabatic potential and the relative phonon frequency. As a consequence, the efficiency of (0200) photoexcitations can be high if the frequency of the bond-coupled phonons is sufficiently higher than that of the charge-coupled phonons. On the other hand, the efficiency of (1010) photoexcitations is insensitive to the relative phonon frequency, though it is lowered if the excitation energy is off-resonant. 

Here, we show that the relative phonon frequency can become an important parameter characterizing the photoinduced melting dynamics if the order is stabilized by different types of phonons. The charge-density oscillation is also affected by the relative phonon frequency, so that theoretical calculations are generally useful for the mode assignment of the coherent oscillation and at least for the identification of interactions stabilizing the order. Furthermore, consideration of relative phonon frequencies may open the possibility for controlling the efficiency of photoinduced dynamics. If many types of phonons are involved with photoinduced dynamics, we conjecture that efficient structural deformation is achieved when the phonon coupled with the initial charge-transfer process has the highest frequency, the secondary phonon induced after the first has the second highest frequency, and so on. 

%\appendix
%\section{}

\begin{acknowledgments}
The authors are grateful to S. Koshihara and K. Onda for showing their data prior to publication and for enlightening discussions.
This work was supported by the Next Generation SuperComputing Project (Nanoscience Program) and Grants-in-Aid for Scientific Research (C) (No. 19540381), for Scientific Research on Priority Area ``Molecular Conductors'' (No. 15073224), and for Creative Scientific Research (No. 15GS0216) from the Ministry of Education, Culture, Sports, Science and Technology, Japan. 
\end{acknowledgments}

\bibliography{pipt07_3}

\end{document}